\documentclass[10pt]{article}\usepackage{longtable}
\usepackage[english]{babel}
\usepackage{amsmath,amsfonts,amssymb,bm}\interdisplaylinepenalty=2500
\ifx\pdfoutput\undefined\usepackage{graphicx}\else\usepackage[pdftex]{graphicx}\fi
\ifx\pdfoutput\undefined\usepackage{rotating}\else\usepackage[pdftex]{rotating}\fi
\graphicspath{{figs/}}
\mathchardef\epsilon="0122   \mathchardef\varepsilon="010F
\mathchardef\theta="123      \mathchardef\vartheta="0112
\mathchardef\rho="125        \mathchardef\varrho="011A
\mathchardef\phi="127        \mathchardef\varphi="011E
\ifx\undefined\degrees\def\degrees{\ensuremath{^{\circ}}}\fi
\ifx\undefined\celsius\def\celsius{\ensuremath{^{\circ}\mathrm{C}}}\fi
\ifx\undefined\unit\def\unit#1{\ensuremath{\mathrm{\,#1}}}\fi
\ifx\undefined\micro\def\micro{\ensuremath{\mu}}\fi
\ifx\undefined\sups\def\sups#1{\ensuremath{^{\mathrm{#1}}}}\fi
\ifx\undefined\subs\def\subs#1{\ensuremath{_{\mathrm{#1}}}}\fi
\ifx\undefined\ohm\def\ohm{\ensuremath{\mathrm{\Omega}}}\fi
\def\req#1{(\ref{#1})}
\def\PrintGraphicFileNeme{1}
\newcommand{\namedgraphics}[2]{
	\parbox{\textwidth}{%
	\ifnum\PrintGraphicFileNeme>0\rotatebox{90}{~\ttfamily\scriptsize#2}\fi%
	\hspace*{\fill}\includegraphics[scale=#1]{#2}\hspace*{\fill}}}

\def\PrintType{3}
\title{On  the \ifnum\PrintType<3\boldmath\fi
$1/f$ Frequency Noise in Ultra-Stable Quartz Oscillators}

\ifnum\PrintType<3
	\author{Enrico~Rubiola, and Vincent Giordano\thanks{%
	The authors are with the FEMTO-ST Institute, UMR\,6174,
	CNRS and Universit\'e de Franche Comt\'e,
	32 av.\ de l'Observatoire, Besan\c{c}on, France.
	e-mail: \{rubiola$|$giordano\}@femto-st.fr}}
	\date{\today}
\else\author{Enrico~Rubiola$^\exists$, and Vincent Giordano\\
	\small $\exists$ web page \texttt{http://rubiola.org}\\[4em]
	\includegraphics[width=0.35\textwidth]{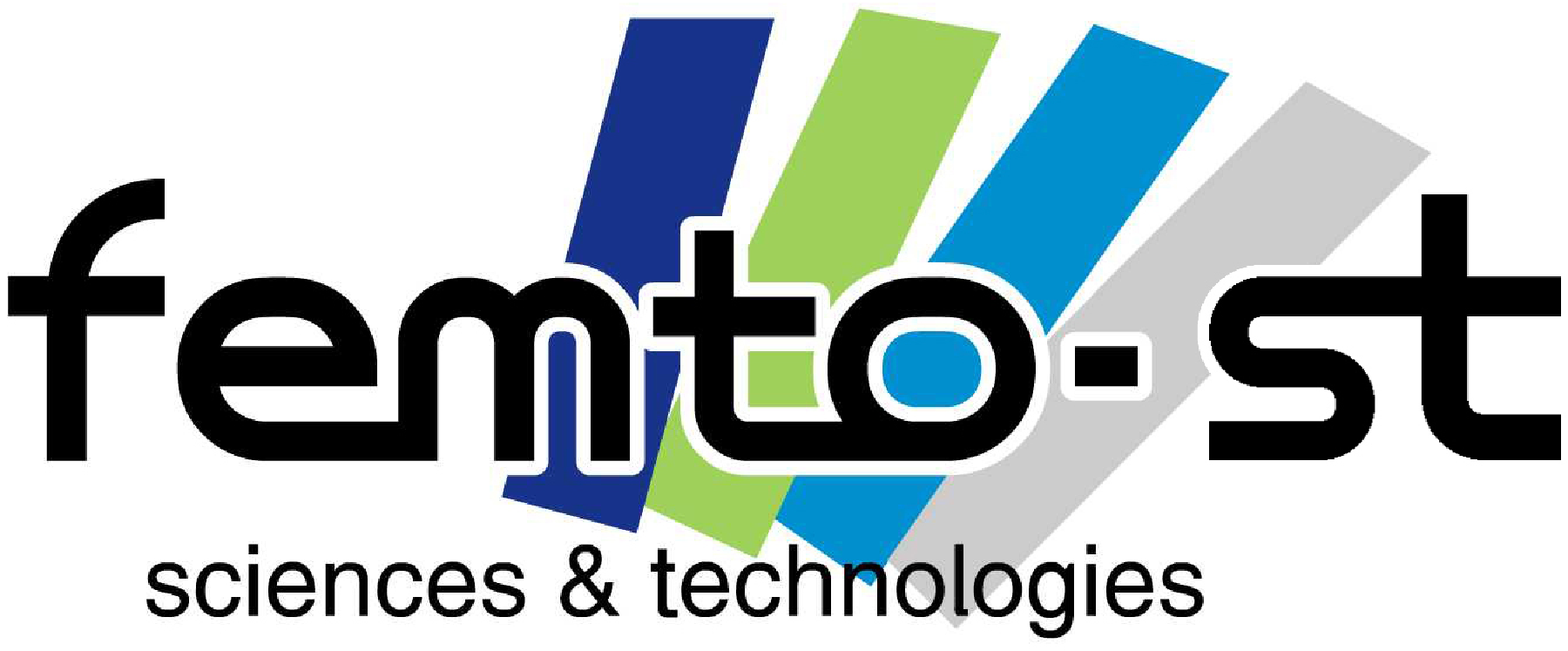}\\[0.5em]
	\small FEMTO-ST Institute\\[-0.5ex]
	\small CNRS and Universit\'e de Franche Comt\'e, 
	\small Besan\c{c}on, France\\[1.5em]}
	\date{\small\today}
	\ifnum\PrintType=3\fi
	\ifnum\PrintType=4\fi

\fi
\begin{document}

\maketitle
\begin{abstract}
\ifnum\PrintType<3\boldmath\fi
The frequency flicker of an oscillator, which appears as a $1/f^3$ line in the phase noise spectral density, and as a floor on the Allan variance plot, originates from two basic phenomena, namely: 
(1) the $1/f$ phase noise turned into $1/f$ frequency noise via the Leeson effect, and 
(2) the $1/f$ fluctuation of the resonator natural frequency.   The discussion on which is the dominant effect, thus on how to improve the stability of the oscillator, has been going on for years without giving a clear answer.  This article tackles the question by analyzing the phase noise spectrum of several commercial oscillators and laboratory prototypes, and demonstrates that the fluctuation of the resonator natural frequency is the dominant effect.  The investigation method starts from reverse engineering the oscillator phase noise in order to show that if the Leeson effect was dominant, the resonator merit factor $Q$ would be too low as compared to the available technology.  
\end{abstract}

\def\SymbolList{%
\begin{center}
\addcontentsline{toc}{section}{Symbol list}
\begin{longtable}{ll}\hline
\multicolumn{2}{l}{\textbf{\large\rule[-1ex]{0pt}{3.5ex}Symbol list}}\\\hline
\rule{0pt}{2.5ex}%
$A$		& amplifier (voltage) gain\\
$b_i$	& coefficient of the power-law representation of $S_\phi(f)$\\
$b(t)$	& resonator impulse response in the phase space\\
$B(s)$	& Laplace transform of $b(t)$\\
$f$		& Fourier frequency (near dc)\\
$f_c$	& amplifier corner frequency, divides white and $1/f$ regions\\
$f_L$	& Leeson frequency\\
$F$		& amplifier noise figure\\
$h_i$	& coefficient of the power-law representation of $S_y(f)$\\
$H(s)$	& loop transfer function (Laplace transform)\\
$\mathcal{H}(s)$	& loop transfer function (Laplace transf.) in the phase space\\
$j$		& imaginary unit, $j^2=-1$\\
$k$		& $1.38{\times}10{-23}$ J/K, Boltzmann constant\\
$P$, $P_0$         & carrier power.  Also $P_a$, $P_b$, $P_m$, etc.\\
$Q$		& resonator merit factor\\
$R$		& frequency-srability ratio\\
$s$		& complex frequency, $s=\sigma+j\omega$, in Laplace transforms\\
$S(f)$, $S_x(f)$ & single-sided power spectrum density (of the quantity $x$)\\
$t$                & time\\
$T$, $T_0$	& absolute temperature, reference temperature ($T_0=290$ K)\\
$v(t)$		& (voltage) signal, as a funtion of time\\
$V_0$, $V_i$	& peak carrier voltage\\
$y$		& fractional frequency fluctuation, $y=\smash{\frac{\nu-\nu_0}{\nu_0}}$\\
$\alpha(t)$	& fractional amplitude fluctuation\\
$\beta(s)$	& resonator impulse response (Laplace transform)\\
$\nu$, $\nu_0$	& frequency, carrier frequency\\
$\sigma_y(\tau_m)$   & Allan deviation of the quantity $y$\\
$\tau_m$	& as in $\sigma_y(\tau_m)$, measurement time\\
$\tau$	&  resonator relaxation time\\
$\phi(t)$	& oscillator phase fluctuation\\
$\Phi(s)$	& Laplace transform of $\phi(t)$\\
$\psi(t)$	& amplifier phase fluctuation\\
$\Psi(s)$	& Laplace transform of $\psi(t)$\\
\rule[-1ex]{0pt}{0ex}%
$\omega$          & angular frequency\\\hline
\multicolumn{2}{l}{\textbf{Note: 
		$\bm{\omega}$ is used as a shorthand for $\bm{2\pi\nu}$
		or $\bm{2\pi f}$, and viceversa}}
\end{longtable}
\end{center}}

\ifnum\PrintType<3%
\else
\clearpage\SymbolList\clearpage\tableofcontents
\fi

\section{Introduction and summary}\label{sec:introduction}
In the domain of ultra-stable quartz oscillators used in the most demanding applications, like space and atomic fountain clocks, we notice that the frequency flicker is often the most critical parameter.  The required stability is sometimes in the upper $10^{-14}$ (Allan deviation) at 1--30 s or so, which can only be achieved in the lower HF band (5--10 MHz), and after selection.  In such cases, identifying the dominant flicker mechanism is far from being trivial.  
Whereas some authors strongly suggest that the amplifier noise can be the parameter that limit the frequency stability, rather than the flickering of the resonator natural frequency \cite{walls95eftf,besson99eftf}, the general literature seems not to give a clear answer.  
This conclusion results from a set of selected articles, which includes the measurement of the frequency stability \cite{walls75im,rubiola00uffc} and the interpretation of the flicker noise of crystal resonators \cite{kroupa88uffc,kroupa05pla};  the design fundamentals of the nowadays BVA resonators \cite{besson77fcs}; some pioneering works on the low-frequency noise in quartz oscillators \cite{brendel75im,driscoll75im}; more recent articles focusing on specific design solutions for ultra-stable oscillators \cite{norton91fcs,norton95fcs,candelier98eftf,candelier03fcs,tuladhar97eftf}; and, as a complement, a thorough review of the SiO$_2$ crystal for the resonator fabrication is found in \cite{brice85rmp}.
Conversely, in everyday-life oscillators, which span from the low-cost XOs to the OCXOs used in telecommunications and instrumentation, the relative simplicity of the low-noise electronics required indicates that the frequency flicker is chiefly the $1/f$ fluctuation of the resonator.   

In a previous work \cite{rubiola05arxiv-leeson}, now extended to more commercial products and laboratory prototypes, we have analyzed the phase noise spectrum of some oscillators, aiming at understanding the internal mechanisms and parameters.   We look at the phase-noise spectrum from the right hand to the left, hence from the higher Fourier frequencies to the lower, matching theory, technology and physical insight.  In this way we get information on the sustaining amplifier on the output buffer, on the Leeson effect and on the resonator. 

In this article we first explain the phase noise mechanisms in amplifiers.  Then we introduce the Leeson effect, which consists of the phase-to-frequency conversion of noise below the resonator cutoff (Leeson) frequency $f_L=\smash{\frac{\nu_0}{2Q}}$.  Finally, we analyze the phase noise spectral density $S_\phi(f)$ of a few oscillators.  The conclusion that the resonator natural frequency is the main cause of frequency flickering is based on experimental facts.  After taking away the effect of the output buffer, we calculate the frequency $f''_L$ at which the oscillator $f^{-3}$ line crosses the $f^{-1}$ line of the sustaining amplifier.  
Provisionally assuming that $f''_L$ is the the Leeson frequency, we observe that the resonator merit factor $Q_s=\smash{\frac{\nu_0}{2f''_L}}$ thereby calculated is far too low for a high-tech resonator.  Conversely, under any reasonable assumption about the true merit factor, the Leeson effect is found at a frequency $f_L\ll f''_L$. Therefore the Leeson $f^{-3}$ line on the $S_\phi(f)$ plot is well hidden below the resonator fluctuation.

\section{Phase noise fundamentals}
Let the quasi-perfect oscillator sinusoidal signal of frequency $\nu_0$
\begin{align}
v(t)=V_0[1+\alpha(t)]\cos[2\pi\nu_0t+\phi(t)]~.
\label{eqn:oscillator-signal}
\end{align}
where $\alpha(t)$ is the fractional amplitude noise, and $\phi(t)$ is the phase noise. 
The AM noise is not essential to this work.  The phase noise is best described in terms of $S_\phi(f)$, i.e., the one-sided power spectral density of $\phi(t)$, as a function of the Fourier frequency $f$.
In addition to $f$, we use the angular frequency $\omega$ for both carrier-related frequencies
($\omega=2\pi\nu$), and Fourier frequencies ($\omega=2\pi f$) without need of introducing it, and the normalized frequency fluctuation $y=\smash{\frac{\nu-\nu_0}{\nu_0}}$.
The quantities $\nu$, $f$ and $y$ refer to one-sided transforms, $\omega$ to two-sided transforms.
Frequency fluctuations are described in terms of $S_y(f)$, related to $S_\phi(f)$ by
\begin{align}
\label{eqn:sy-sphi}
S_y(f)=\frac{f^2}{\nu_0^2}\,S_\phi(f)~.
\end{align}

A model that has been found useful in describing the oscillator noise
spectra is the power-law
\begin{align}
\label{eqn:power-law-sphi-sy}
S_y(f)=\sum_{i=-2}^{2}h_if^i \quad\Leftrightarrow\quad
S_\phi(f)=\sum_{i=-4}^{0}b_if^i~.
\end{align}
Our main concern is the frequency flickering term $b_{-3}f^{-3}$, which is related to the Allan variance by
\begin{align}
\label{eqn:allan-flicker}
\sigma^2_y   \;=\; 2\ln(2) \: h_{-1}
			\;=\; 2\ln(2) \: \frac{b_{-3}}{\nu_0^2}~, 
\end{align}
constant, i.e., independent of the measurement time.

Finally, the general background on phase noise and frequency stability is available from numerous references, among which we prefer \cite{rutman78pieee}, \cite{ccir90rep580-3}, \cite{kroupa:frequency-stability}, and \cite[Vol.\,1, Chapter 2]{vanier:frequency-standards}. A IEEE standard is also available \cite{ieee99std1139}.

\section{Phase noise in rf (and microwave) amplifiers}\label{sec:amplifier-noise}
\begin{figure}[t]
\ifnum\PrintType<3\centering\includegraphics[scale=0.8]{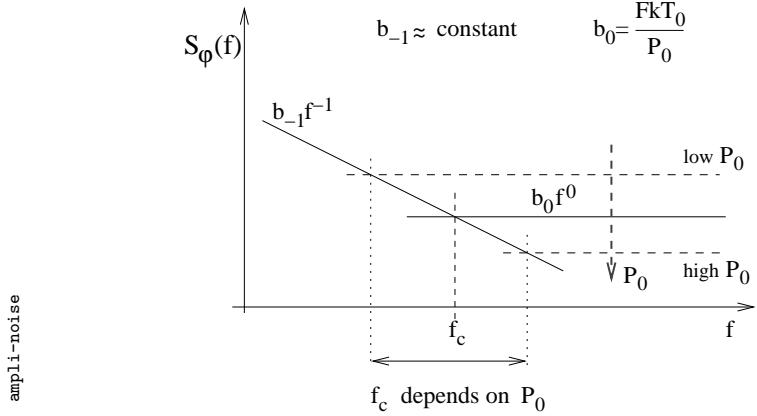}
	\else\namedgraphics{0.8}{ampli-noise}\fi
\caption{Typical phase noise spectral density of an amplifier.}
\label{fig:ampli-noise}
\end{figure}
\subsection{White noise}\label{ssec:amplifier-white}
The equivalent noise spectrum density at the amplifier input is $N=FkT_0$, where $F$ is the noise figure and $kT_0$ is the thermal energy.  This type of noise is additive.  In the presence of a carrier of power $P_0$, the phase noise spectral density is 
\begin{align}
\label{eqn:ampli-white} 
S_\phi(f)=b_0f^0~\text{(constant)}
\end{align}
with
\begin{align}
\label{eqn:ampli-b0} 
b_0=\frac{FkT_0}{P_0}~.
\end{align}
When amplifiers are cascaded, the noise contribution of each stage is divided by the gain of all the preceding stages (Friis formula \cite{friis44ire}).  Accordingly, in most practical cases the total noise is chiefly the noise of the first stage.  Of course, this also holds for phase noise.  

\subsection{Flicker noise}\label{ssec:amplifier-flicker}
Understanding the close-in noise starts from the bare observation that the output spectrum is of the white type---flat in a wide frequency range---when the carrier power is zero, and that noise shows up close to the carrier only when a sufficiently large carrier power is present at the amplifier output. 
The obvious consequence is that the close-in flickering results from a parametric effect by which the near-dc flicker noise modulates the carrier in amplitude and phase.

The simplest way to understand the noise up-conversion is to model the amplifier signal as a nonlinear function truncated to the 2nd order
\begin{align}
v_o(t) = a_1v_i(t) + a_2v_i^2(t) + \ldots~,
\label{eqn:nonlinear-ampli} 
\end{align}
in which the complex input signal
\begin{align}
v_i(t) = V_i \, e^{j\omega_0t} + n'(t) + jn''(t)
\label{eqn:carrier-noise} 
\end{align}
contains the carrier and the internally generated near-dc noise $n(t)=n'(t)+jn''(t)$. Rather than being an easy-to-identify voltage or current, $n(t)$ is an abstract random signal that also accounts for the efficiency of the modulation process.  Combining \req{eqn:nonlinear-ampli} and \req{eqn:carrier-noise} and selecting the terms close to the carrier frequency $\omega_0$, we get 
\begin{align}
v_o(t) &= V_i \Bigl\{ a_1+2a_2\bigl[n'(t) + jn''(t)\bigr] \Bigr\} e^{j\omega_0t}~.
\label{eqn:vout-analyt} 
\end{align}
Hence, the random fluctuations are
\begin{align}
\alpha(t)=2\:\frac{a_2}{a_1}\:n'(t) \quad\text{and}\quad
\phi(t)=2\:\frac{a_2}{a_1}\:n''(t)~.
\label{eqn:ampli-random-phase} 
\end{align}

Deriving Eq.~\req{eqn:ampli-random-phase}, the statistical properties of $n'(t)$ and $n''(t)$ are not affected by the carrier power.  This accounts for the experimental observation that the amplifier phase noise given in \unit{rad^2/Hz} is about independent of power in a wide range \cite{halford68fcs,walls97uffc,hati03fcs}.  
Thus
\begin{align}
S_\phi(f)=b_{-1}f^{-1} \qquad\text{$b_{-1}\approx$ constant}~.
\label{eqn:ampli-flicker} 
\end{align}
Of course, some dependence on $P_0$ remains.  We ascribe it to terms of order higher than 2 in \req{eqn:nonlinear-ampli}, and to the effect of the large signal regime on the dc bias.  
In the case of bipolar amplifiers used in HF/VHF amplifiers, $b_{-1}$ is in the range of $10^{-12}$ to $10^{-14}$ \unit{rad^2/Hz} ($-120$ to $-140$ \unit{dBrad^2/Hz}). 

When $m$ amplifiers are cascaded, the The Friis formula does not apply.  Instead, the phase noise barely adds 
\begin{align}
\label{eqn:add-flicker}
(b_{-1})_\text{cascade} = \sum_{i=1}^m (b_{-1})_i~.
\end{align}
This occurs because the $1/f$ phase noise is about independent of power.  Of course, the amplifiers are supposed independent.

\subsection{Phase noise spectrum}
Combining white noise [Eq.~\req{eqn:ampli-white}] and flicker noise [Eq.~\req{eqn:ampli-flicker}], there results the spectral density $S_\phi(f)$ shown in Fig.~\ref{fig:ampli-noise}. 
It is important to understand that the white noise term $b_{0}f^{0}$ depends on the carrier power $P_0$, while the flicker term $b_{-1}f^{-1}$ does not.  Accordingly, the corner frequency $f_c$ at which $b_{-1}f^{-1}=b_0$ is a function of $P_0$, thus $f_c$ should \emph{not} be used to describe noise.  The parameters $b_{-1}$, $F$, and $P_0$ should be used instead.

\section{Phase noise in feedback oscillators}\label{sec:oscillator-noise}
\subsection{The Leeson effect}\label{ssec:leeson}
\begin{figure}[t]
\ifnum\PrintType<3\centering\includegraphics[scale=0.8]{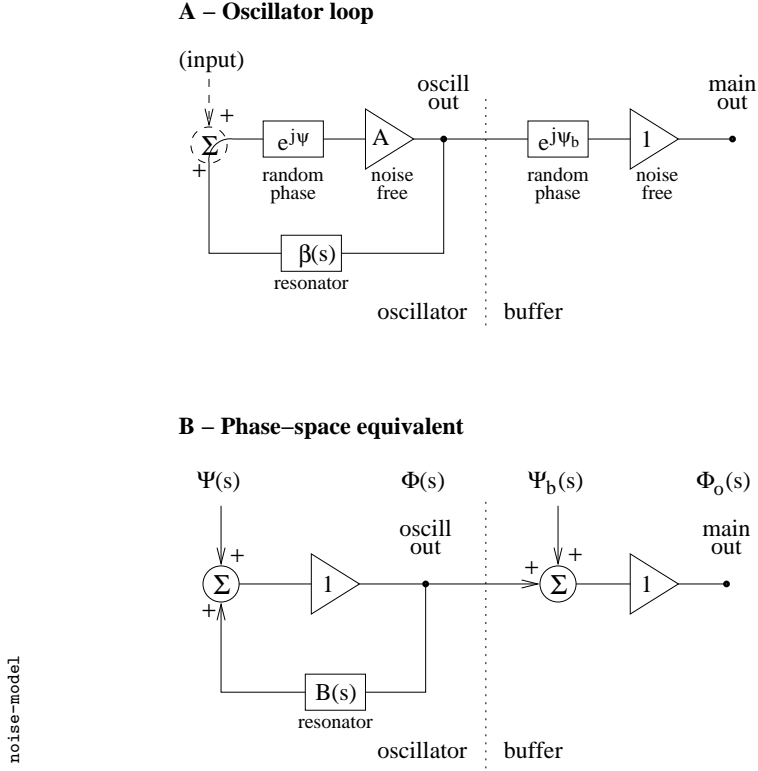}
	\else\namedgraphics{0.80}{noise-model}\fi
\caption{Oscillator model and its phase-space equivalent.  For the sake of simplicity all the dependence on $s$ is moved to $\beta(s)$, hence the gain $A$ is assumed constant.  The scheme emphasizes the amplifier phase noise.  Amplitude noise is not considered.}
\label{fig:noise-model}
\end{figure}
Figure~\ref{fig:noise-model} shows a model for the feedback oscillator, and its equivalent in the phase space. 
All signals are the Laplace transform of the time-domain quantities, as a function of the complex frequency $s=\sigma+j\omega$. 
The oscillator transfer function is derived from Fig.~\ref{fig:noise-model}\,A  according to the basic rules of linear systems 
\begin{align}
\label{eqn:xfer}
H(s)	 \;=\; \frac{1}{\beta(s)}\:\frac{1}{\dfrac{1}{A\beta(s)}-1}
	 \;=\; \frac{A}{1-A\beta(s)}
\end{align}
Stationary oscillation occurs at the angular frequency $\omega_0$ at which $A\beta(j\omega)=1$, thus $|A\beta(j\omega)|=1$ and $\arg[A\beta(j\omega)]=0$.  This is known as the Barkhausen condition for oscillation.  At $s=j\omega_0$ the denominator of $H(s)$ is zero, hence oscillation is sustained with zero input signal.  
Oscillation starts from noise or from the switch-on transient if $\Re\{A\beta(s)|_{s=j\omega_0}\}>1$ (yet only slightly greater than $1$ for practical reasons).  When the oscillation reaches a threshold amplitude, the loop gain is reduced to $1$ by saturation.  The excess power is pushed into harmonics multiple of $\omega_0$, and blocked by the resonator.   For this reason, at $\omega_0$ the oscillator operates in \emph{quasi-linear} regime.

In most quartz oscillators, the sustaining amplifier takes the form of a negative resistance that compensates for the resonator loss.  Such negative resistance is interpreted (and implemented) as a transconductance amplifier that senses the voltage across the input and feeds a current back to it.  Therefore, the negative-resistance oscillator loop is fully equivalent to that shown in Fig.~\ref{fig:noise-model}.

In 1966, D. B. Leeson \cite{leeson66pieee} suggested that the oscillator phase noise is described by
\begin{align}
\label{eqn:leeson}
S_\phi(f)=\left[1+\frac{1}{f^2}\,\frac{\nu_0^2}{4Q^2}\right]S_\psi(f) \qquad\text{(Leeson)}\,,
\end{align}
This formula calls for the phase-space representation of Fig.~\ref{fig:noise-model}\,B, which deserves the following comments.

The Laplace transform of the phase of a sinusoid is probably the most common mathematical tool in the domain of PLLs \cite{klapper-frankle:pll,gardner:pll,best:pll,egan:pll}.  Yet it is unusual in the analysis of oscillators.  
The phase-space representation is interesting in that \emph{the phase noise turns into additive noise, and the system becomes linear}. The noise-free amplifier barely repeats the input phase, for it shows a gain exactly equal to one, with no error. The resonator transfer function, i.e., the Laplace transform of the impulse response, is
\begin{align}
\label{eqn:resonator-bigb}
B(s)=\frac{1}{1+s\tau}\qquad \tau=\frac{2Q}{\omega_0}~.
\end{align}
The inverse time constant is the low-pass cutoff angular frequency of the resonator
\begin{align}
\label{eqn:leeson-omega}
\omega_L=\frac1\tau=\frac{\omega_0}{2Q}~.
\end{align}
The corresponding frequency
\begin{align}
\label{eqn:leeson-frequency}
f_L \;=\; \frac{\omega_L}{2\pi} \;=\; \frac{1}{2\pi\tau} \;=\; \frac{\nu_0}{2Q}
\end{align}
is known as the Leeson frequency. 
Equation~\req{eqn:resonator-bigb} is proved in two steps: 
\begin{enumerate} 
\item Feed a Heaviside step function $\kappa U(t)$ in the argument of the resonator input sinusoid.  The latter becomes $\cos\left[\omega_0t+\kappa U(t)\right]$.
\item Linearize the system for $\kappa\rightarrow0$.  
This is correct in low phase noise conditions, which is certainly our case. 
Accordingly, the input signal becomes $\cos(\omega_0t)-\kappa\sin(\omega_0t)U(t)$.
\item Calculate the Laplace transform of the step response, and use the property that the Laplace transform maps the time-domain derivative into a multiplication by the complex frequency $s$. The Dirac function $\delta(t)$ is the derivative of $U(t)$.  
\end{enumerate}
The full mathematical details of the proof are available in \cite[Chapter 3]{rubiola05arxiv-leeson}. 

Applying the basic rules of linear systems to Fig.~\ref{fig:noise-model}\,B, we find the transfer function
\begin{align}
\label{eqn:mathcal-h}
\mathcal{H}(s) \;=\; \frac{\Phi(s)}{\Psi(s)} 
			\;=\; \frac{1}{1-B(s)} \;=\;	\frac{1+s\tau}{s\tau}~,
\end{align}
thus
\begin{align}
\label{eqn:h2jomega}
\left|\mathcal{H}(j\omega)\right|^2=\frac{1+\omega^2\tau^2}{\omega^2\tau^2}~.
\end{align}
The Leeson formula \req{eqn:leeson} derives from Eq.~\req{eqn:h2jomega} by replacing
\begin{align}
\label{eqn:}
\omega=2\pi f	\quad\text{and}\quad	\tau=\frac{Q}{\pi\nu_0}~.
\end{align}

The transfer function $\mathcal{H}(s)$ has a pole in the origin (pure integrator), which explains the Leeson effect, i.e., the phase-to-frequency noise conversion at low Fourier frequencies.
At high Fourier frequencies it holds that $\mathcal{H}(j\omega)=1$.  In this region, the oscillator noise is barely the noise of the sustaining amplifier.

\begin{figure}[t]
\ifnum\PrintType<3\centering\includegraphics[scale=0.8]{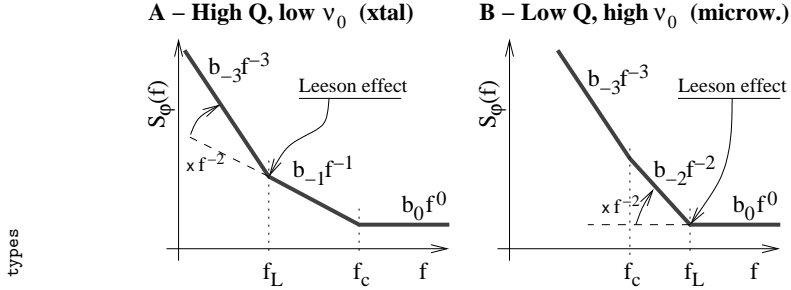}
	\else\namedgraphics{0.8}{types}\fi
\caption{Oscillator phase noise spectra, not accounting for the output buffer.}
\label{fig:types}
\end{figure}
The amplifier phase noise spectrum contains flicker and white noise, i.e., $S_\psi(f)=\left(b_{-1}\right)_\text{ampli}f^{-1}+\left(b_0\right)_\text{ampli}$.  Feeding such $S_\psi(f)$ into the Leeson formula \req{eqn:leeson}, the oscillator $S_\phi(f)$ can only be one of those shown in Fig.~\ref{fig:types}.
Denoting with $f_c$ the corner frequency at which flicker noise equals white noise, 
we often find $f_L<f_c$ in HF/VHF high-$Q$ oscillators, and $f_L>f_c$ in microwave oscillators.  In ultra-stable HF quartz oscillators (5--10 MHz), the spectrum is always of the type A ($f_L<f_c$).

\subsection{Output buffer} 
The phase noise $S_{\psi\,b}(f)$ of the output buffer barely adds to the oscillator phase noise
\begin{align}
\label{eqn:leeson-w-buffer}
S_{\phi\,o}(f)=\left[1+\frac{1}{f^2}\,\frac{\nu_0^2}{4Q^2}\right]S_\psi(f)+S_{\psi\,b}(f)~.
\end{align}
This a consequence of the flicker noise mechanism explained Section~\ref{ssec:amplifier-flicker}, and
inherent in the model of Fig.~\ref{fig:noise-model}\,B.

\subsection{Resonator stability}
The oscillator frequency follows the random fluctuation of the resonator natural frequency.  However complex or tedious the formal proof for this statement can be, the experimentalist is familiar with the fact that the quartz oscillator can be frequency-modulated by a signal of frequency far higher than the Leeson frequency.  For example, a 5 MHz oscillator based on a $Q=2{\times}10^6$ resonator shows a Leeson frequency of $1.25$ Hz (see Table~\ref{tab:parameters}), while it can be modulated by a signal in the kHz region.  
Additionally, as a matter of fact, the modulation index does not change law from below to beyond the Leeson frequency.  This occurs because the modulation input acts on a varactor in series to the quartz, whose capacitance is a part of the motional parameters.

\subsection{Other effects}\label{ssec:other-effects}
The sustaining amplifier of a quartz oscillator always includes some kind of feedback; often the feedback is used to implement a negative resistance that makes the resonator oscillate by nulling its internal resistance.
The input admittance $Y_i$ seen at the amplifier input can be represented as 
\begin{align}
Y_i = Y_i^{(v)} + Y_i^{(r)}~,
\end{align}
that is, the sum of a virtual term $(v)$ plus a real term $(r)$.  
The difference between `virtual' and `real' is that in the case of the virtual admittance the input current flows into the feedback path, while in the case of the real admittance the input current flows through a grounded dipole.  This is exactly the same concept of virtual impedance routinely used in the domain of analog circuits \cite[Chapter~1]{franco:operational-amplifiers}. 
The admittance $\smash{Y_i^{(r)}}$ also includes the the effect of the pulling capacitance in series to the resonator, and the stray capacitances of the electrical layout.
As a consequence, the fluctuation $\delta\smash{Y_i^{(v)}}$ is already accounted for in the amplifier noise, hence in the model of Fig.~\ref{fig:noise-model}, while the fluctuation $\delta\smash{Y_i^{(r)}}$ is not.
On the other hand, $\smash{Y_i^{(r)}}$ interacts with the resonator parameters, thus $\delta\smash{Y_i^{(r)}}$
yields frequency fluctuations not included in the Leeson effect.
The hard assumption is made in our analysis, that $|\delta\smash{Y_i^{(r)}}|^2\ll|\delta\smash{Y_i^{(v)}}|^2$.
In words, we assume that the fluctuation of the electronics are chiefly due to the gain mechanism of the amplifier.  Whereas the variety of circuits is such that we can not provide a proof for this hypothesis, common sense suggests that electronics works in this way.

\def\TableOne{%
\begin{tabular}{lcccccccccccl}\hline
\ifnum\PrintType>2\\[-1em]\hline\fi
\rule[-2ex]{0pt}{5.0ex}%
Oscillator & $\nu_0$ &
\makebox[0pt]{$(b_{-3})_\text{tot}$} & \makebox[0pt]{$(b_{-1})_\text{tot}$} 
	& \makebox[0pt]{$(b_{-1})_\text{amp}$} &
$f'_L$ & $f''_L$ & $Q_s$ & $Q_t$ & 
$f_L$ & $(b_{-3})_\text{L}$ & $R$ & \parbox{6ex}{\centering Ref.\,\&\\[-0.3ex]Note}\\\hline
\ifnum\PrintType>2\\[-1em]\hline\fi
\rule[-2ex]{0pt}{5.5ex}%
\parbox{10ex}{Oscilloquartz\\[-0.3ex]8600} & $5$ &
$-124.0$ & $-131.0$ & $-137.0$ &
\makebox[0pt]{$2.24$} & $4.5$ & $5.6{\times}10^5$ & \makebox[0pt]{$1.8{\times}10^6$} & 
$1.4$ & $-134.1$ & $10.1$ & 
	\parbox{13ex}{\mbox{\cite{tuladhar97eftf,www.oscilloquartz.com}}\\[-0.3ex](1)}\\  
\ifnum\PrintType>2\hline\fi
\rule[-2.0ex]{0pt}{5.5ex}%
\parbox{11ex}{Oscilloquartz\\[-0.3ex]8607} & $5$ &
$-128.5$ & $-132.5$ & $-138.5$ &
$1.6$    & $3.2$    & $7.9{\times}10^5$ & $2{\times}10^6$ & 
\makebox[0pt]{$1.25$} & $-136.5$ & $8.1$ &
	\parbox{13ex}{\mbox{\cite{tuladhar97eftf,www.oscilloquartz.com}}\\[-0.3ex](1)}\\  
\ifnum\PrintType>2\hline\fi
\rule[-2.0ex]{0pt}{5.5ex}%
\parbox{11ex}{CMAC\\[-0.3ex]Pharao} & $5$ &
$-132.0$ & $-135.5$ & $-141.1$ &
$1.5$    & $3$    & $8.4{\times}10^5$ & $2{\times}10^6$ & 
\makebox[0pt]{$1.25$}   & $-139.6$ & $7.6$ &
	\parbox{13ex}{\mbox{\cite{candelier98eftf,candelier03fcs,www.cmac.com}}\\[-0.3ex](1)}\\  
\ifnum\PrintType>2\hline\fi
\rule[-2.0ex]{0pt}{5.5ex}%
\parbox{11ex}{\mbox{FEMTO-ST}\\[-0.3ex]\mbox{LD protot.}} & $10$ &
$-116.6$ & $-130.0$ & $-136.0$ &
$4.7$    & $9.3$    & $5.4{\times}10^5$ & \makebox[0pt]{$1.15{\times}10^6$} & 
$4.3$ & $-123.2$ & $6.6$ &
	\parbox{13ex}{\mbox{\cite{galliou04fcs}}\\[-0.3ex](3)}\\  
\ifnum\PrintType>2\hline\fi
\rule[-2.0ex]{0pt}{5.5ex}%
\parbox{11ex}{Agilent\\[-0.3ex]10811} & $10$ &
$-103.0$ & $-131.0$ & $-137.0$ &
$25$    & $50$    & $1{\times}10^5$ & $7{\times}10^5$ & 
$7.1$ & $-119.9$ & $16.9$ &
	\parbox{13ex}{\mbox{\cite{burgoon81hpj}}\\[-0.3ex](4)}\\  
\ifnum\PrintType>2\hline\fi
\rule[-2.0ex]{0pt}{5.5ex}%
\parbox{11ex}{Agilent\\[-0.3ex]prototype} & $10$ &
$-102.0$ & $-126.0$ & $-132.0$ &
$16$    & $32$    & $1.6{\times}10^5$ & $7{\times}10^5$ & 
$7.1$ & $-114.9$ & $12.9$ &
	\parbox{13ex}{\mbox{\cite{karlquist00uffc}}\\[-0.3ex](5)}\\  
\ifnum\PrintType>2\hline\fi
\rule[-2.5ex]{0pt}{5.5ex}%
\parbox{11ex}{Wenzel\\[-0.3ex]\mbox{501-04623}} & $100$ &
$-67.0$ & $-132$\,? & $-138$\,? &
\makebox[0pt]{$1800$}    & \makebox[0pt]{$3500$}    
				& $1.4{\times}10^4$ & $8{\times}10^4$ & 
\makebox[0pt]{$625$} & $-79.1$ & $15.1$ &
	\parbox{13ex}{\mbox{\cite{www.wenzel.com}}\\[-0.3ex](3)}\\\hline  
\ifnum\PrintType>2\\[-1em]\hline\fi
\rule[-2.2ex]{0pt}{6ex}%
unit     & MHz &
\parbox{8ex}{\centering dB\\\unit{rad^2/Hz}} & 
\parbox{8ex}{\centering dB\\\unit{rad^2/Hz}} & 
\parbox{8ex}{\centering dB\\\unit{rad^2/Hz}} & 
Hz & Hz & (none) & (none) & Hz & \parbox{8ex}{\centering dB\\\unit{rad^2/Hz}} & dB & \\\hline
\ifnum\PrintType>2\\[-1em]\hline\fi
\multicolumn{13}{l}{\rule[0ex]{0pt}{2.5ex}Notes}\\
\multicolumn{13}{l}{(1) Data are from specifications, full options about low noise and high stability.}\\
\multicolumn{13}{l}{(2) Measured by CMAC on a sample.  CMAC confirmed that 
$\smash{2{\times}10^6<Q<2.2{\times}10^6}$ in actual conditions.  See Fig.~\ref{fig:candelier}.} \\ 
\multicolumn{13}{l}{(3) LD cut, built and measured in our laboratory, yet by a different team.  All design parameters are known, hence $Q_t$.}\\
\multicolumn{13}{l}{(4) Measured by Hewlett Packard (now Agilent) on a sample.}\\
\multicolumn{13}{l}{(5) Implements a bridge scheme for the degeneration of the amplifier noise.
  Same resonator of the Agilent 10811.}\\
\multicolumn{13}{l}{\rule[-1ex]{0pt}{0.5ex}(6) Data are from specifications. See Fig.~\ref{fig:wenzel}.}\\\hline
\ifnum\PrintType>2\\[-1em]\hline\fi
\end{tabular}}

\begin{table*}[t]
\ifnum\PrintType<3%
	\caption{Estimated Parameters of some Ultra-Stable Oscillators.}\label{tab:parameters}
	\centering\TableOne
\else\begin{sideways}\begin{minipage}{0.96\textheight}
       \vspace*{0ex}
	\caption{Estimated Parameters of some Ultra-Stable Oscillators.\hspace*{15ex}}\label{tab:parameters}
	\setlength{\unitlength}{1ex}
	\begin{picture}(0,0)
	\put(-14,-36){\mbox{\TableOne}}%
	\end{picture}
	\end{minipage}\end{sideways}\fi
\end{table*}

\section{Analysis of the oscillator phase noise}\label{sec:analysis}
\begin{figure}[t]
\ifnum\PrintType<3\centering\includegraphics[scale=0.8]{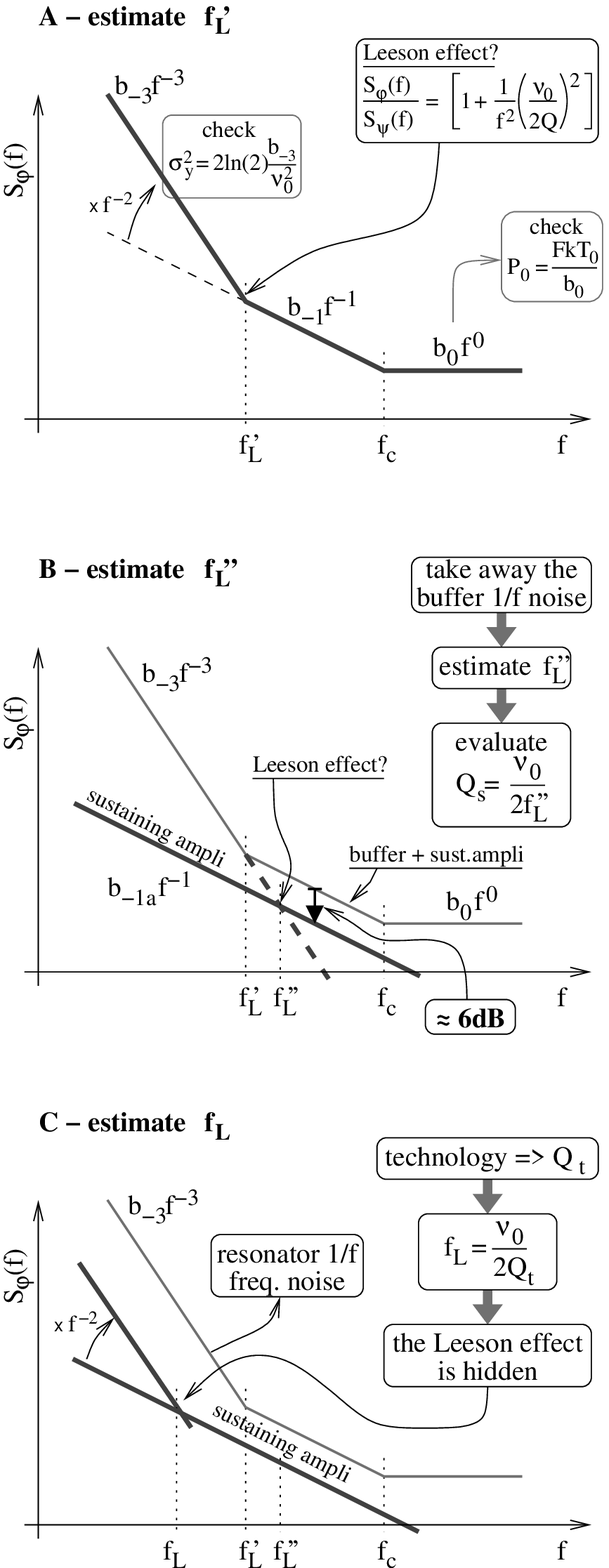}
	\else\namedgraphics{0.79}{interpretation}\fi
\caption{Interpretation of the phase noise in quartz oscillators.}
\label{fig:interpretation}
\end{figure}
This section address the core question, wether the $1/f^3$ noise observed on the oscillator $S_\phi(f)$ plot is due to the Leeson effect, or it originates in the resonator.  The interpretation method is shown in Fig.~\ref{fig:interpretation}, and discussed below.

\begin{enumerate}
\item 
We start from the spectrum, measured or taken from the oscillator specifications.  The first step is to remove the residual of the mains (50 or 60 Hz and multiples) and other stray signals, and to fit the spectrum with the power-law [Eq.~\req{eqn:power-law-sphi-sy}].
This process is called parametric estimation of the spectrum.  With a pinch of experience, sliding old-fashion squares on a A4-size plot gives unexpectedly good results.  Otherwise, the mathematical methods explained in \cite{jenkins:spectral-analysis,percival:spectral-analysis} are useful.
After this, the spectrum looks like that of Fig.~\ref{fig:interpretation}\,A\@, 

\item
The term $b_0f^0$ is chiefly due to the sustaining amplifier, hence the amplifier input power can be calculated using Eq.~\req{eqn:ampli-b0}
\begin{align}
P_0=\frac{FkT_0}{b_0}~.
\end{align}
In the absence of information, it is wise to take $F=1.26$ (1 dB).
To the extent of our analysis, estimating $P_0$ is only a check of plausibility.

\item
Feeding the oscillator $b_{-3}$ term into Eq.~\req{eqn:allan-flicker}, we calculate the floor of the Allan deviation $\sigma_y$.  We check on the consistency between calculated value and specifications or measures, if available.

\item
At first sight, the cutoff frequency $f'_L$ (Fig.~\ref{fig:interpretation}\,A) can be taken for the Leeson frequency because there the slope changes from $f^{-3}$ to $f^{-1}$.
Yet the term $b_{-1}f^{-1}$ contains the flicker of the sustaining amplifier and of the output buffer, which add [Equations~\req{eqn:add-flicker} and \req{eqn:leeson-w-buffer}].  For this reason, $f'_L$ can not be the Leeson frequency.

\item
Actual oscillators have 2--4 buffer stages, the main purpose of which is to isolate the feedback loop from the environment in order to ensure frequency stability and to prevent injection locking.  Owing to the Leeson effect, a wise designer will spend the lowest-noise technology in the sustaining amplifier, rather than in the buffer.  Thus, we assume that the buffer contributes 3/4 of the total noise, and that sustaining amplifier contributes 1/4 ($-6$ dB). 
Accordingly, we plot the line $b_{-1\,a}\,f^{-1}$ in Fig.~\ref{fig:interpretation}\,B, 6 dB below the total flicker.

\item
After taking away the buffer noise, the continuation of the $b_{-3}f^{-3}$ line meets the $b_{-1\,a}f^{-1}$ line at $f=f''_L$.  The latter is a new candidate for the Leeson frequency.  
Feeding $f''_L$ into Eq.~\req{eqn:leeson-frequency}, we calculate the resonator merit factor $Q_s$
(the subscript $s$ stands for `spectrum')
\begin{align}
Q_s=\frac{\nu_0}{2f''_L}~.
\end{align}

\item
Technology suggests a merit factor $Q_t$ (the subscript $t$ stands for `technology') significantly larger than $Q_s$, even in actual load conditions.
Feeding $Q_t$ into Eq.~\req{eqn:leeson-frequency}, we calculate $f_L$ based on the actual merit factor
\begin{align}
f_L=\frac{\nu_0}{2Q_t}~,  
\end{align}
as shown in Fig.~\ref{fig:interpretation}\,C\@.  There follows a phase noise term $(b_{-3})_L$, which account for the Leeson effect alone.

\item
Given $Q_t\gg Q_s$, thus $f_L \ll f''_L$, the Leeson effect is hidden.
Consequently, the oscillator $f^{-3}$ phase noise is chiefly due to the fluctuation of the resonator natural frequency. 

\end{enumerate}
We introduce the stability ratio $R$, defined as
\begin{align}
R=\frac{(\sigma_y)_\text{oscill}}{(\sigma_y)_\text{Leeson}} \qquad\text{(floor)},
\end{align}
and related to the other oscillator parameters by
\begin{align}
\label{eqn:r-q}
R	\;=\; \sqrt{\frac{(b_{-3})_\text{tot}}{(b_{-3})_L}} 
	\;=\; \frac{Q_t}{Q_s} \;=\; \frac{f''_L}{f_L} ~.
\end{align}
This can be demonstrated from the $b_{-3}$ term of the Leeson formula \req{eqn:leeson}, using Equations~\req{eqn:allan-flicker} and \req{eqn:leeson-frequency}.
The parameter $R$ states how bad is the actual oscillator, as compared to the same oscillator governed only by the Leeson effect, with the resonator fluctuations removed.  
Thus, $R=1$ (0 dB) indicates that the oscillator $f^{-3}$ phase noise comes from the Leeson effect.  Equal contribution of resonator and Leeson effect yield $R=\sqrt{2}$ (3 dB), while $R\gg\sqrt{2}$ is found when resonator instability is the main cause of $f^{-3}$ phase noise.
In all cases we have analyzed, discussed in the next Section, we find $R$  of the order of 10 dB, with a minimum of 6.6 dB\@.  This means that the Leeson effect is hidden below the frequency fluctuation of the resonator.

Coming back to the estimation of the $1/f$ noise of the sustaining amplifier it is to be remarked that if the $1/f$ noise of this is lower than 1/4 of the total flicker, $f''_L$ is further pushed on the right hand on Fig.~\ref{fig:interpretation}\,B-C, which reinforces the conclusion that the resonator is the main cause of frequency fluctuation.

\section{Experimental data and discussion}\label{sec:examples}
\begin{figure}[t]
\ifnum\PrintType<3\centering\includegraphics[scale=0.55]{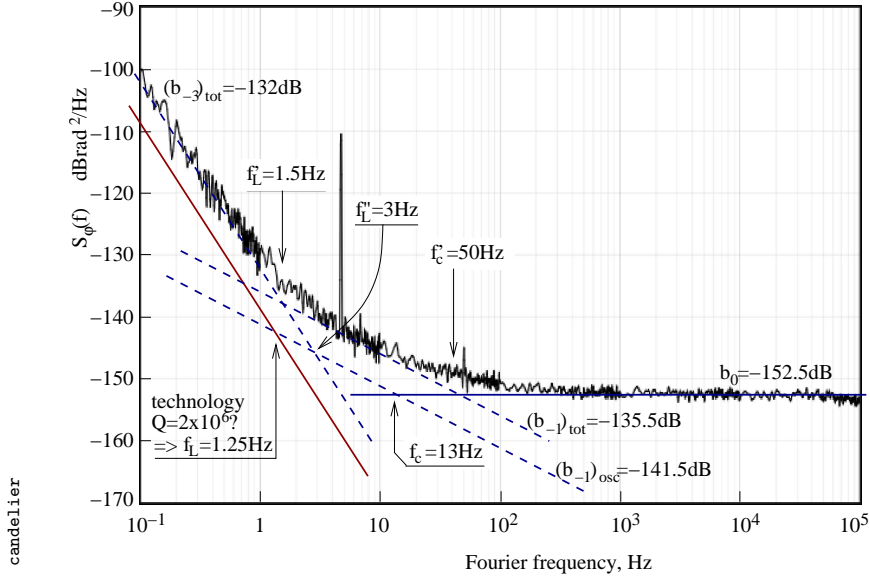}
	\else\namedgraphics{0.70}{candelier}\fi
\caption{Phase noise of the CMAC Pharao 5 MHz quartz OCXO.  Courtesy of CMAC\@.  Interpretation and mistakes are of the authors.}
\label{fig:candelier}
\end{figure}
Figure~\ref{fig:candelier} shows the phase noise spectrum of a 5 MHz oscillator, out of a small series intended as the flywheel for the space Cesium fountain clock Pharao \cite{lemonde01:cold-atoms,bize04crphysique}.  On this plot, the reader can follow the interpretation process explained in Section~\ref{sec:analysis}, and illustrated in Fig.~\ref{fig:interpretation}.  Guessing on technology, the merit factor was estimated to be $2{\times}10^{6}$.  Afterwards, the manufacturer confirmed \cite{candelier:personal-comm} that $Q_t$ is between $2{\times}10^{6}$ and $2.2{\times}10^{6}$ in actual load conditions for that series of oscillators, and that the flicker noise of the sustaining amplifier is less than 1/4 ($-6$ dB) of the total flickering.  This validates our conclusions. 

Table \ref{tab:parameters} shows the results of our analysis on some oscillators.  
The ability to estimate the resonator merit factor is necessary to understand the oscillator inside.  
Experience indicates that the product $\nu_0Q$ is a technical constant of the piezoelectric quartz resonator, in the range from $1{\times}10^{13}$ to  $2{\times}10^{13}$.  As a matter of fact, the highest values are found in the 5 MHz resonators. 
In load conditions, the resonator merit factor is somewhat lower.  The actual value depends on frequency, on the designer skill, and on the budget for implementation.
A bunch of data are available from \cite{gerber-ballato:frequency-control,kroupa05pla,walls95eftf}, and from our early attempts to measure the resonator frequency stability \cite{rubiola00uffc}.  The oscillators we have analyzed exhibit the highest available stability, for we are confident about published data.  The Agilent 10811 (hence the Agilent prototype) is closer to the routine production, and probably closer to the cost-performance tradeoff, as compared to the other ones, thus understanding oscillator the inside is more difficult. 
Nonetheless, in this case the value of $Q_s$ is so low that there is no doubt that it can not be the resonator merit factor.

In the case of the Oscilloquartz 8607, the $f^{-3}$ noise is too low for it to be extracted from the $S_\phi(f)$ spectrum available on data sheet, which starts from 1 Hz. 
Yet, we can use the device specifications $\smash{S_\phi(f)|_{1\unit{Hz}}}=-127$ \unit{dBrad^2/Hz}, $\smash{S_\phi(f)|_{10\unit{Hz}}}=-142$ \unit{dBrad^2/Hz}, and $\smash{S_\phi(f)|_{1\unit{kHz}}}=-153$ \unit{dBrad^2/Hz}.  In fact, looking at the spectrum and at the Allan variance it is clear that at $f=1$ Hz and $f=10$ Hz the terms $b_{-3}f^{-3}$ and $b_{-1}f^{-1}$ determine $S_\phi(f)$, with at most a minor contribution of $b_0$. 
It is also clear that $\smash{S_\phi(f)|_{1\unit{kHz}}}\simeq b_0$.
Thus $b_{-3}$ and $b_{-1}$ are obtained by solving a system of two equations like $S_\phi(f)=b_{-3}f^{-3}+b_{-1}f^{-1}+b_0$, at 1 Hz and 10 Hz.

\begin{figure}[t]
\ifnum\PrintType<3\centering\includegraphics[scale=0.7]{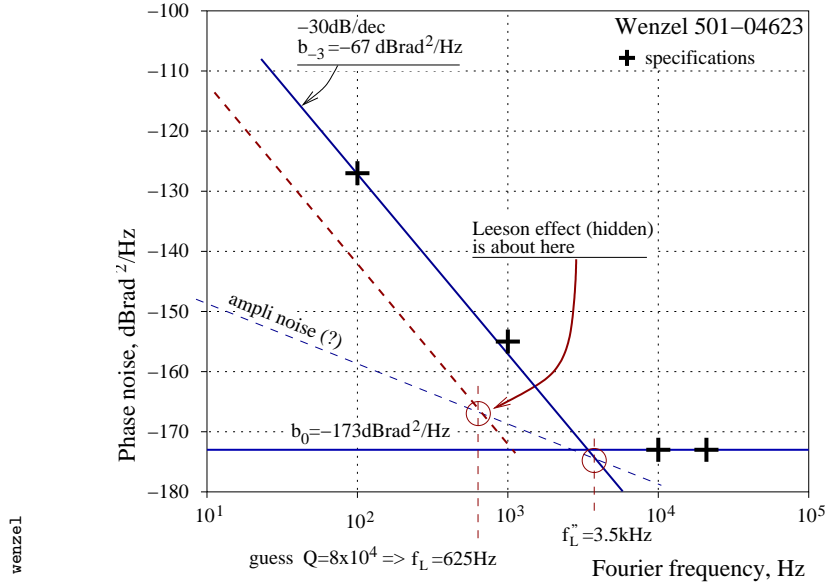}
	\else\namedgraphics{0.8}{wenzel}\fi
\caption{Phase noise of the Wenzel 501-04623 100 MHz quartz OCXO \cite{www.wenzel.com}.}
\label{fig:wenzel}
\end{figure}
In the case of the Wenzel 501-04623 oscillator (Fig.~\ref{fig:wenzel}), the specifications available on the manufacturer web site consist of a few points, while the whole spectrum is not published.  Experience indicates that in the case of 100 MHz oscillators the $f^{-1}$ line tends to be hidden by the frequency flickering.  That said, we can only guess that the $f^{-1}$ noise of the sustaining amplifier is similar to that of other oscillators.  This is sufficient to estimate $f''_L$, and to notice that  the merit factor $Q_s$ is far too low as compared to the state of the technology, and to conclude that the $f^{-3}$ phase noise is due to the fluctuation of the resonator natural frequency. 
It is to be remarked that the power at the amplifier input is of the order of 10--20 $\mu$W in all other cases, and of 1 mW here.  In addition, the 100 MHz resonator is smaller in size than the other resonator.  A relatively high frequency flicker is therefore not surprising.

The examples shown above indicate that, under the assumption of Sections~\ref{sec:amplifier-noise}--\ref{sec:oscillator-noise}, the oscillator frequency flickering is chiefly due to the fluctuation of the resonator natural frequency.

\section*{Acknowledgements}\addcontentsline{toc}{section}{Acknowledgements}
We are indebted to Jean-Pierre Aubry (Oscilloquartz, Switzerland), and Vincent Candelier (CMAC, France) for providing spectra and support.  
Jacques Grolambert (FEMTO-ST, retired) set up the methods and the knowledge on oscillator measurement in our laboratory.
R\'emi Brendel (FEMTO-ST), Giorgio Brida (INRIM, Italy), Lute Maleki and G. John Dick (JPL, USA)  helped us with several discussions.  R\'emi Brendel has taken a personal interest in our work and has offered a wealth of suggestions and constructive criticism; we owe him special thanks.

\def\bibfile#1{/Users/rubiola/Documents/work/bib/#1}
\ifnum\PrintType<3\bibliographystyle{ieeetr}
\else\bibliographystyle{amsalpha}\addcontentsline{toc}{section}{References}\fi
\bibliography{\bibfile{ref-short},\bibfile{references},\bibfile{rubiola}}
\end{document}